\documentclass[twocolumn,pra,english,showkeys,superscriptaddress,longbibliography]{revtex4-1}
\usepackage[utf8x]{inputenc}
\usepackage{amsfonts}
\usepackage{amsmath}
\usepackage{amssymb}
\usepackage[colorlinks, citecolor=blue,linkcolor=red]{hyperref}
\usepackage{color}
\usepackage{float}
\usepackage{hyperref}
\usepackage{textcomp}
\usepackage[rightcaption]{sidecap}
\usepackage{subfigure}
\usepackage[rightcaption]{sidecap}
\usepackage{graphicx}
\begin{document}
\title{Engineering chaos in a four-mirror cavity-optomechanics with mechanical drives}
\author{Kashif Ammar Yasir}
\email{kayasir@zjnu.edu.cn}\affiliation{Department of Physics, Zhejiang Normal University, Jinhua 321004, China.}
\author{Gao Xianlong}
\email{gaoxl@zjnu.edu.cn}
\affiliation{Department of Physics, Zhejiang Normal University, Jinhua 321004, China.}
\setlength{\parskip}{0pt}
\setlength{\belowcaptionskip}{-10pt}
\begin{abstract}
We study occurrence of chaos in a four-mirror optomechanical cavity with mechanical drives externally interacting with two transversely located moving-end mirrors of the cavity. The strong cavity mode, driven by the pump laser, excites mechanical oscillations in both moving-end mirrors with its radiation pressure. These radiation-pressure-induced mechanical effects then lead to the indirect coupling between two transverse mirrors, where intra-cavity field mimics as a spring between two mechanical objects. By computing Poincar\'e surface of sections for both mirrors over a wide interval of initial conditions, we illustrate the transition from stable to mixed -- containing stable islands and chaotic seas -- Poincar\'e surface of sections with external mechanical drives. To further explore the occurrence of chaos with mechanical drives, we measure the spatio-temporal responses of moving-end mirrors initially located in mixed Poincar\'e sections. We find that both of the mirrors follow chaotic temporal evolution with external mechanical drives, even in the absence of any one of the mechanical drives. To quantitatively measure the occurrence of chaos, we computed the possible Lyapunov exponents and collective Kolmogorov-Sinai Entropy of the system. We find that the largest Lyapunov exponent, and corresponding Kolmogorov-Sinai Entropy, not only gains positive values with increase in external drives but also crucially depends on the initial conditions chosen from the Poincar\'e surface of sections. Furthermore, we show the enhancement in chaotic dynamics of mirrors in the presence of mechanical damping rates associated with the oscillatory motion of the mirrors.     
\end{abstract}
\date{\today}
\maketitle

\section{Introduction}
Optomechanics -- a manifestation of radiation pressure to produce mechanics in micro/macroscopic resonators -- has emerged as a fascinating subject, especially with respect to cavity quantum electrodynamics, and is the subject of increasing investigations \cite{Kippenberg08,Kipp09,Meystre13}. The demonstration of mechanical characteristics of light in macroscopic regime has enabled to design gravitational wave detector \cite{gw1,gw2} and, in microscopic regime, has led us to perform highly accurate measurements \cite{Rugar} and develop optomechanical crystals \cite{EchenfieldNat2012}. Optomechanics has provided us a better opportunity to demonstrate and examine the ground state cooling of quantum mirrors \cite{CornnellNat2010,teufel2011,chan2011,Steele2015,Arcizet2006,Gigan2006} and quantum nonlinear optical interaction leading to the optomechanically induced transparencies \cite{Ref17,Ref8,Ref9}. Further, the coupling of multiple mechanical objects, notably ultra-cold atoms \cite{Esslinger}, with optomechanical system leads to the multi-species hybrid cavity-optomechanics. The hybridization of cavity-optomechanics yields in the demonstrations of ultra-cold atomic induced quantum cooling \cite{peter2}, many-body quantum entanglement \cite{Vitali2012,Vitali2014,Sete2014,Hofer}, high fidelity state transfer \cite{YingPRL2012,SinghPRL2012}, multiple electromagnetically induced transparencies \cite{agarwal2010,Stefan2010,Peng2014,SafaviNaeini2011}, and cavity-optomechanics with synthetically dressed atomic states \cite{kashif4,kashif5,kashif55}. 

The stable and unstable dynamics of resonators are crucial to develop complete picture of an optomechanical system, especially in hybrid environment. This motivation has led researchers to the study of optical \cite{kashif3,kashif33,Yang2011} as well as spatial bistabilities \cite{Meystre2010} in hybrid optomechanics and to the stunning demonstration of chaos in cavity-optomechanics induced by the nonlinear interaction of radiation pressure \cite{chaos1,chaos2,chaos3,chaos4,chaos5,chaos6,chaos7}. The nonlinear interactions of radiation pressure to engineer chaos in optomechanics \cite{Hilborn}, further yield in the demonstration of dynamical localization -- a quantum mechanical phenomena emerging from the maximum quantum interference in chaotic domain \cite{Saif2005} -- for both optomechanical mirror \cite{kashif1} and ultra-cold atoms \cite{kashif2}. The recent investigations on four-mirror optomechanical system, with two transversely located moving-end mirrors, have provided a new setup to study hybrid and complex system \cite{Ref283,Ref284,Ref285,Ref286}. The impacts of external mechanical drives on the mechanical resonators of such system have led to multiple novel investigations \cite{Ref281,Ref282}. However, a study on the dynamical effects of mechanical drive, especially in regards of chaos, in four-mirror optomechanical system is needed. Further, the discussion on the dynamical aspects such as chaos and disorder in a transversely location two-body system -- where the coupling between two-body is generated with the intra-cavity photonic radiation pressure -- is very significant to understand multiple chaos mediated quantum phenomena.    

In this paper, we discuss chaos in a four-mirror cavity engineered with two mechanical drives externally interacting with two moving-end mirrors, which are transversely located along $x$-axis and $y$-axis of the cavity. The beam splitter (BS), located at the center of the cavity, splits the cavity mode which then gets coupled with the moving-end mirrors. The mechanical characteristics of light in the form of radiation pressure excite mechanical oscillations in the moving-end mirrors. These mechanical oscillations contain bistable behavior due to the nonlinear interaction of the cavity radiation pressure. The unstable state among the two stable states of bistability leads to the chaotic evolution of the mirrors at strong mirror-field couplings, as it has been studied for the conventional optomechanical systems by coupling multiple oscillators with cavity \cite{kashif3,Meystre2010,chaos1,chaos2}. But, in our study, we use external mechanical drives in a four-mirror cavity setup to produce chaotic response in mechanical oscillators even at weak mirror-cavity coupling, which makes it novel not only in a setup but also in the approach to obtain chaos. 
\begin{figure}[tp]
	\includegraphics[width=7cm]{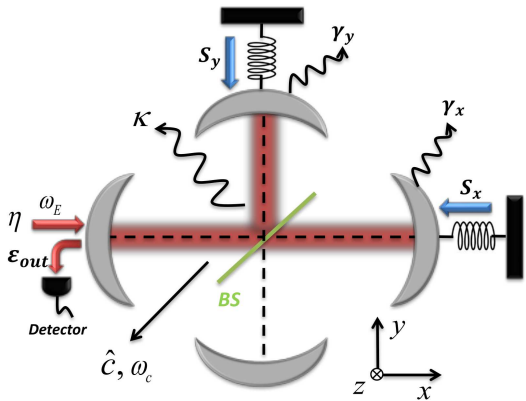}
	\caption{The systematic sketch of an optomechanical four-mirror cavity two transversely ($x$-axis and $y$-axis) coupled moving-end mirrors. A pump laser longitudinally drives the cavity while two lasers for mechanical drives -- $S_x$ along $x$-axis and $S_y$ along $y$-axis -- externally interact with mirrors. $\eta$ and $\omega_E$ are the intensity and frequency of pump laser, respectively. The cavity mode, excited by the pump laser, engineers coupling between transverse moving-end mirrors of the cavity after splitting from beam splitter (BS) located at the center of the cavity.}
	\label{fig1}
\end{figure}

By plotting Poincar\'e surface of sections for the possible interval of initial conditions, we show that the presence of external mechanical drive leads to the transformation of the regular Poincar\'e sections to the mixed Poincar\'e sections, containing both stable islands and chaotic seas, for both moving-end mirrors. The mechanical drive of one mirror also imprints chaotic signatures in the second mirror even in the absence of mechanical drive for that mirror. To further strengthen our argument, we illustrate the emergence of chaos by calculating the temporal evolution of the mirrors by choosing their initial states from the Poincar\'e sections. We also illustrate the effects of initial conditions on the chaotic temporal evolution. To perform quantitative analysis of chaos, we calculated the possible Lyapunov exponents and collective Kolmogorov-Sinai Entropy of the system. We find that the largest Lyapunov exponent moves to positive domain with increase in external drives, which indeed yields in positive Kolmogorov-Sinai Entropy. But both Lyapunov exponents and Kolmogorov-Sinai Entropy is appeared to be sensitive to the initial position chosen from the Poincar\'e surface of sections. Which means, if the system is not initially in chaotic sea, then the increase in external drives will less rapidly enhance the Lyapunov exponents and Kolmogorov-Sinai Entropy as compare to when the system initially in chaotic sea. Furthermore, we discuss the influences of mechanical damping on the chaotic behavior and find that the increase in mechanical damping rate enhances the chaotic behavior. 

The contents of the manuscript are distributed as follow. Section \ref{sec1} accommodates the system modeling and mathematical details. Section \ref{sec2} contains the results and discussion on Poincar\'e surface of sections. Section \ref{sec3} describes the spatio-temporal dynamics of moving-end mirrors with respect to the external mechanical drive. Section \ref{sec4} demonstrates quantitative analysis of the occurrence of chaos with external mechanical drives by illustrating Lyapunov exponents and Kolmogorov-Sinai Entropy. Section \ref{sec5} contains the influences of mechanical damping on the chaotic dynamics. Finally, the conclusion is given in section \ref{sec6}. 

\section{System description and Hamiltonian}\label{sec1}
We consider a four-mirror cavity-optomechanics consisting of two moving-end mirrors located along $x$-axis and $y$-axis of the cavity, as illustrated in Fig. \ref{fig1}. The strong cavity mode, driven by the external pump laser having intensity $\eta$ and frequency $\omega_E$, gets split from the partial BS and interacts with the moving-end mirrors of the cavity. The radiation pressure exerted by the cavity mode excites multistable vibrations in both of the mirrors, similarly as it does in the conventional optomechanical systems \cite{kashif3,Meystre2010}. In this study, however, we use two mechanical drives -- externally interacting and perturbing the mechanical motion of the moving-end mirrors -- to engineer Hamiltonian chaos for the moving-end mirrors of the cavity.  

The system Hamiltonian, containing the energies associated with both of the moving-end mirrors, intra-cavity field and their mutual coupling, can be expressed as \cite{kashif1,kashif3,cklaw},
\begin{eqnarray}
\hat{H} &=& \sum_{j=x,y}\bigg(\frac{\hbar\omega_j}{2}\big(\hat{p}^2_j+\hat{q}^2_j\big)+\hbar G_j\hat{c}^{\dag}\hat{c}\hat{q}_j \nonumber \\
&&+S_j q_j cos(\delta_j t+\phi_j)\bigg)+\hbar\Delta_{c}\hat{c}^{\dag}\hat{c}\nonumber \\
&&-i\hbar\eta(\hat{c}-\hat{c}^{\dag}).\label{h1}
\end{eqnarray}
Here the first term corresponds to the motion of moving-end mirror defined with the dimensionless position quadrature $\hat{q}_j=(\hat{a}_{j}+\hat{a}^{\dag}_{j})/\sqrt{2}$ and momentum quadrature $\hat{p}_j=i(\hat{a}_{j}-\hat{a}^{\dag}_{j})/\sqrt{2}$, with subscript $j=x,y$ corresponding to the mirrors along $x$- and $y$-axis, respectively. These quadrature obey the canonical commutation relation $[\hat{q}_j,\hat{p}_j]=i$ and are defined over the phononic annihilation and creation operators $\hat{a}_{j}$ and $\hat{a}^{\dag}_{j}$, respectively \cite{kashif5,cklaw}. The second term appearing in the Hamiltonian defines the coupling between moving-end mirrors and cavity mode resulting because of radiation pressure. $\hat{c}$ ($\hat{c}^\dag$) corresponds to the photonic annihilation (creation) field operator for the cavity mode while $G_j=\sqrt{2}(\omega_{c}/L_j)x_{j}, (j=x,y)$ defines the coupling strength with zero point spatial oscillations $x_{j}=\sqrt{\hbar/2m\omega_{j}}$ for mirror mass $m$.  Here $\omega_{j}=\omega_x, \omega_y$ is the frequency of moving-end mirrors oriented along cavity arm along $x$-axis and $y$-axis, respectively, with length $L_j=L_{x,y}$. We assume that both of the moving-end mirrors are similar and have the same mass $m$. Further, the lengths of transverse arms (longitudinal and vertical) are same, $i.e. L_x=L_y$.  

The third term in Hamiltonian (\ref{h1}) accommodates the coupling between external mechanical drives and the moving-end mirrors, defined with the mechanical force strength $S_j=\mathcal{\alpha}_j\sqrt{(\hbar/\omega_j m_j)}, (j=x,y)$. Here $\mathcal{\alpha}_j=\mathcal{\alpha}_x, \mathcal{\alpha}_y$ corresponds to the amplitudes while $\delta_j=\delta_{x,y}$ and $\phi_j=\phi_{x,y}$ are the frequency and the phase of the mechanical force interacting with moving-end mirrors along $x$-axis and $y$-axis, respectively \cite{Ref283,Ref284,Ref285,Ref286,Ref281,Ref282}. The fourth term defines the energies of cavity mode with respect to the cavity pump detuning $\Delta_c=\omega_E-\omega_c$, where $\omega_c$ is the frequency of cavity mode. The last term defines the coupling between cavity mode and external pump laser, where $\vert\eta\vert=\sqrt{P\times\kappa/\hbar\omega_{E}}$ is the coupling strength parameter.

To incorporate the effects of damping and noises associated with mechanical motion of moving-end mirrors and cavity mode, we use Heisenberg Langevin equation (QLE) approach to govern spatio-temporal behavior for each degree of freedom associated with system. The QLEs with consideration of standard noise operators and damping will be read as,
\begin{eqnarray}\label{eq4}
\frac{d\hat{c}}{dt}&=&(i\Delta_c+\sum_{j=x,y}iG_{j}\hat{q}_j-\kappa)\hat{c}+\eta +\sqrt{2\kappa} \hat{c}_{in},\label{2a} \\ 
\frac{d\hat{q}_j}{dt}&=&\omega_{j}\hat{p}_j, (j=x,y)\label{2d}\\
\frac{d\hat{p}_j}{dt}&=&-\omega_{j}\hat{q}_j+G_{j}\hat{c}^{\dag}\hat{c}
+S_j cos (\delta_j t+\phi_j)  \nonumber \\
&&-\gamma_{j}\hat{p_j}+\hat{F}_{j}, (j=x,y).\label{2e}
\end{eqnarray}
Here $\kappa$ corresponds to the effective intra-cavity field decay rate, including the photon leakage (or scattering) towards the bottom mirror, oriented along $-x$-axis, from BS. $\hat{c}_{in}$ represents noises yielding from cavity input and defined by the Markovian noise operator with zero-mean $\langle \hat{c}_{in}(t)\rangle=0$ and temporal delta-correlation $\langle \hat{c}_{in}(t)\hat{c}_{in}^{\dagger}(\acute{t})\rangle=\delta(t-\acute{t})$. $\gamma_j=\gamma_{x,y}$ accommodates the mechanical damping rates associated with moving-end mirror vibrating along $x$-axis and $y$-axis, respectively. $\hat{F}_{j=x,y}$ is the zero-mean Langevin-force operator corresponding to the quantum noises associated with the Brownian motion of moving-end mirrors along $x$- and $y$-axis. These Langevin-force operators can be defined with non-Markovian correlation $\langle \hat{F}_j(t)\hat{F}_j(\acute{t})\rangle=\frac{\gamma_j}{2\pi\omega_j}\int d\omega e^{-i\omega(t-\acute{t})}[1+coth(\frac{\hbar\omega}{2k_BT})], (j=x,y)$ \cite{Dalibard2011,Pater06}, where $k_B$ is the Boltzmann constant and $T$ is the temperature of thermal reservoir around the system. However, in this work, by adopting strong cavity mode regime (i.e. $\hbar\omega_c>>k_BT$) and, respectively considering, higher oscillations in strong coupling regime $\hbar\omega_j>>k_BT, (j=x,y)$, we ignore the effects of these noises.

By using above mentioned approximation and treating intra-cavity field as classical variable, one can extract the steady-state value of cavity field (i.e. $\frac{d\hat{c}}{dt}\rightarrow 0$) from QLEs,
\begin{eqnarray}
c_{s}&=&\frac{\eta}{\kappa -i(\Delta_{c}+\sum_{j=x,y}G_{j}q_j)}.
\end{eqnarray}\label{equ2}
Here, $c_s$ corresponds to the steady-state cavity field. Further, by substituting the steady-state value of cavity field into the QLEs for moving-end mirrors, (\ref{2d}) and (\ref{2e}), and ignoring the influences of associated noises, one can derive the equation of motion (EoM) for both of the moving-end mirrors as,
\begin{eqnarray}
&&\frac{1}{\omega_{j}}\frac{d^{2} q_j}{dt^{2}}+\frac{\gamma_{j}}{\omega_{j}}\frac{dq_j}{dt}+\omega_{j} q_j
-\frac{G_j\eta^{2}}{\kappa^{2}
	+(\Delta_c+\sum_{i=x,y}G_{i}q_{i})^2}\nonumber \\
&& =S_j Cos(\delta_j t+\phi_j), (j=x,y).\label{eqmo}
\end{eqnarray}
The EoM possesses crucial importance in order to govern the spatio-temporal dynamics for both moving-end mirrors. Further, from EoM one can derive the effective Hamiltonian in semi-classical domain $\hat{H}_{eff}=E+V$, where the kinetic energies for both of the moving-end mirrors can be simply written as $E=\sum_{j=x,y} (\hbar\omega_{j}/2) p_j^2$. On the other hand, the potential energies corresponding to the dynamical oscillation for both moving-end mirrors can be extracted from the second-order derivative term of EoM \cite{Meystre2010,kashif1,kashif2},
\begin{eqnarray}
V&=&\sum_{j=x,y}\int \frac{d^2 q_j}{dt^2} dq_j\nonumber\\
&=&\sum_{j=x,y}\bigg[-\frac{\omega_{j} q^2_j}{2}+S_j q_j cos(\delta_j t+\phi_j) \nonumber \\
&& +\frac{\eta^{2}}{\kappa}
arctan\big[(\Delta_c+\sum_{i=x,y}G_{i}q_{i})/\kappa\big]\bigg].\label{effV}
\end{eqnarray}
In order to govern Hamiltonian dynamics over the spatio-momentum space, we adopted Hamilton's equation approach on the effective Hamiltonian of the system, 
\begin{eqnarray}
\frac{d p_j}{dt}=-\frac{\partial \hat{H}_{eff}}{\partial q_j}, \frac{dq_j}{dt}=+\frac{\partial \hat{H}_{eff}}{\partial p_j}, (j=x,y). \label{hh}
\end{eqnarray}
On the other hand, one can directly use EoM to obtain the system dynamics. But, to make it a generalized, we adopted Hamilton's equation approach. 
\begin{figure*}[htp]
	\includegraphics[width=17cm]{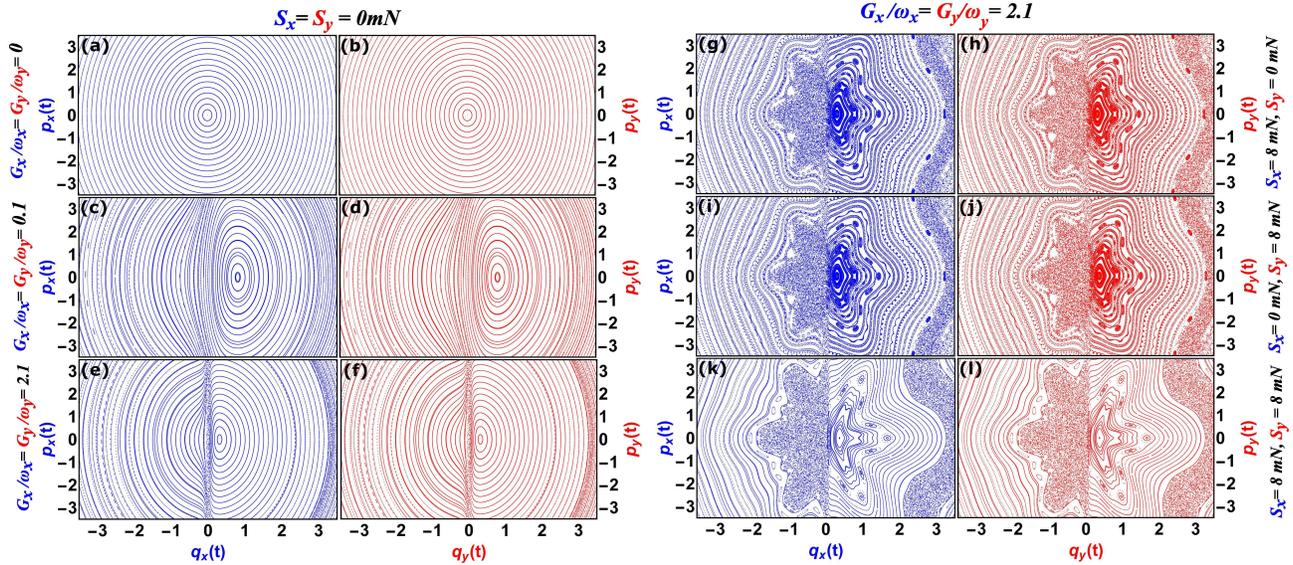}
	\caption{Poincar\'e surface of sections for both of the moving-end mirrors $q_x,p_x$ and $q_y,p_y$ oriented along $x$-axis and $y$-axis. (a), (c) and (e) [(b), (d) and (f)] correspond to the cavity-mirror coupling $G_x=0 \omega_x$, $0.1\omega_x$ and $2.1\omega_x$ [$G_y=0 \omega_y$, $0.1\omega_y$ and $2.1\omega_y$], respectively, for mirror along $x$-axis [$y$-axis] at $S_x=S_y=0$. While (g), (i) and (k) [(h), (j) and (l)] illustrate the effects of external mechanical drive when $S_x=8 mN$, $0 mN$ and $8 mN$ [$S_y=0 mN$, $8 mN$ and $8 mN$], respectively, at $G_x/\omega_x=G_y/\omega_y=2.1$. Here the damping rates for the mirrors are considered zero $\gamma_x=\gamma_y=0$. The other system parameters are considered as $\omega_x=\omega_y=0.1\Delta_c$, and $\kappa\approx1.3\times2\pi$MHz.}
	\label{fig2}
\end{figure*}

To make our findings experimentally possible, we adopted particular set of parameters that are available in recent state-of-the-art experiments \cite{Kippenberg08,Refex1,Refex2}. We consider a four-mirror cavity with two transverse arms with length $L=1.25\times10^{-4}$ and driven by a single mode pump laser, with power $P=0.0164mW$ and frequency $\omega_{E}=3.8\times2\pi\times10^{14}Hz$. The pump field produces a strong cavity mode, with frequency $\omega_{c}$ almost equal to the external pump laser for the sake of quantum nonlinear interactions. The collective decay rate is $\kappa\approx 1\times2\pi MHz$. The cavity field equally exerts radiation pressure on both mirror generating frequency $\omega_{j}= 15.2\times2\pi kHz$, with $j=x,y$ corresponds to the condensate located along $x$-axis and $y$-axis, respectively. The coupling between the intra-cavity field and the mirrors $G_j$, with $j=x,y$, can be defined as $G_j/\omega_j= 1.1\times2\pi MHz$, which we assume to be equal for both arms under condition $G_j/\omega_j>>\kappa$, with $j=x,y$. However, we modulate the effective coupling strengths $G_{x,y}$ in order to obtain the desire results. 

\section{Mixed Poincar\'e surface of sections}\label{sec2}
Dynamics of any mechanical system can be fully understood by its phase space because it contains all possible spatio-momentum (or with any other conjugate variables) states of the system at any time depending upon the initial conditions. The phase space can be obtained by visualizing spatio-momentum temporal evolution on the two-dimensional plane by taking the time-lapse snapshot over the temporal dimension. A slice of spatio-momentum plan at a particular time corresponds to Poincar\'e surface of sections (or Poincar\'e sections) and each point on Poincar\'e sections represents the recurrence of spatio-momentum trajectory. Such spatio-momentum dynamics can conventionally be governed from the Hamilton's equations (\ref{hh}) of the system. For regular or stable system, the spatio-momentum trajectories will follow distinct, without any interference, patterns on the phase space (mostly in the form of circles) depending upon the initial condition. Whereas, in the case of unstable or perturbed system, these trajectories will interfere with each other and can result in chaos, in the form of random seas, in Poincar\'e sections. The dependence of spatio-momentum trajectories on the initial condition (initial values of position and momentum) is crucial. For instance, some initial conditions may lead spatio-momentum trajectories to take regular forms, but some other initial conditions may take the same system towards chaos. Such phase space portraits result to mixed Poincar\'e sections, containing both stable islands and chaotic seas \cite{refc}.  

In order to govern spatio-momentum phase space dynamics, we numerically solve the Hamilton's equations (\ref{hh}), simultaneously for both of the mirrors with respect to three-dimensional quadratic space $(q_j,p_j,q_{j0})$, with $j=x,y$. Here $q_{j0}$ corresponds to the initial condition (or initial position of the moving-end mirrors) for each trajectory, which we considered normally distributed over the interval $q_{j0}\rightarrow[-2\pi,2\pi]$. We consider both of the moving-end mirrors initially at rest, yielding to $p_{j0}\rightarrow 0$, in order to exactly map initial cavity-length configuration on the Poincar\'e sections. For each initial condition, we numerically evolve the system to $t\rightarrow 800\pi$ and plot the Poincar\'e sections, by measuring the recurrence over the $2\pi$ time interval, with respect to the external mechanical drives $S_x$ and $S_y$ for both of the moving-end mirrors, as illustrated in Fig. \ref{fig2}.

In absence of mechanical drives $S_x=S_y=0$ and coupling with the cavity $G_x=G_y=0$, both of the mirrors follow a regular (or stable) Poincar\'e sections with distinct circular trajectories -- without facing any interference -- as illustrated in Figs. \ref{fig2}(a) and \ref{fig2}(b) for moving-end mirrors oriented along $x$-axis and $y$-axis, respectively. Here each circular trajectory corresponds to each initial condition following a symmetric behavior around $q_{x}=q_{y}=0$. However, when we apply a weak coupling between intra-cavity field and the moving-end mirrors, a bend (or tilt) starts appearing in the phase space trajectories asymmetrically squeezing the circular patterns of Poincar\'e sections towards the center at $q_{x}=q_{y}=0$, as can be seen in Figs. \ref{fig2}(c) and \ref{fig2}(d), where $G_x=G_y=0.1$. If we further increase the strength of cavity-mirror couplings, the bend in the phase space trajectories further increase resulting in more asymmetric behavior, see Figs. \ref{fig2}(e) and \ref{fig2}(f), where $G_x=G_y=2.1$.

The effects of coupling between moving-end mirrors and cavity field on the Poincar\'e surface of sections can be understood by analogically modeling cavity as a spring connecting both moving-end mirrors, even though both mirrors are not along same axis. The modification in the spring effects of the cavity field acts as a perturbation to the mechanical oscillations of the mirrors. In other words, increase in the mirror-cavity couplings, due to the increase in radiation pressure, yields to unstable potential energies for the mechanical motion of the moving-end mirrors. These spring effects (or the mirror-cavity couplings) will show more influences when the mirrors move away from the origin $q_x=q_y=0$ along negative $x$-axis and $y$-axis (or when the cavity length decreases with motion of the mirrors). This will result in more squeezing or compression for the circular trajectories on the left-half plane of Poincar\'e section as compared to the right-half plane leading to the asymmetric phase space, see Figs. \ref{fig2}(e) and \ref{fig2}(f). The compression of circular trajectories consequently result in the overlap of these trajectories around $q_x=q_y=0$. But still there are no significant signatures for the chaos.  

\begin{figure*}[tp]
	\includegraphics[width=17cm]{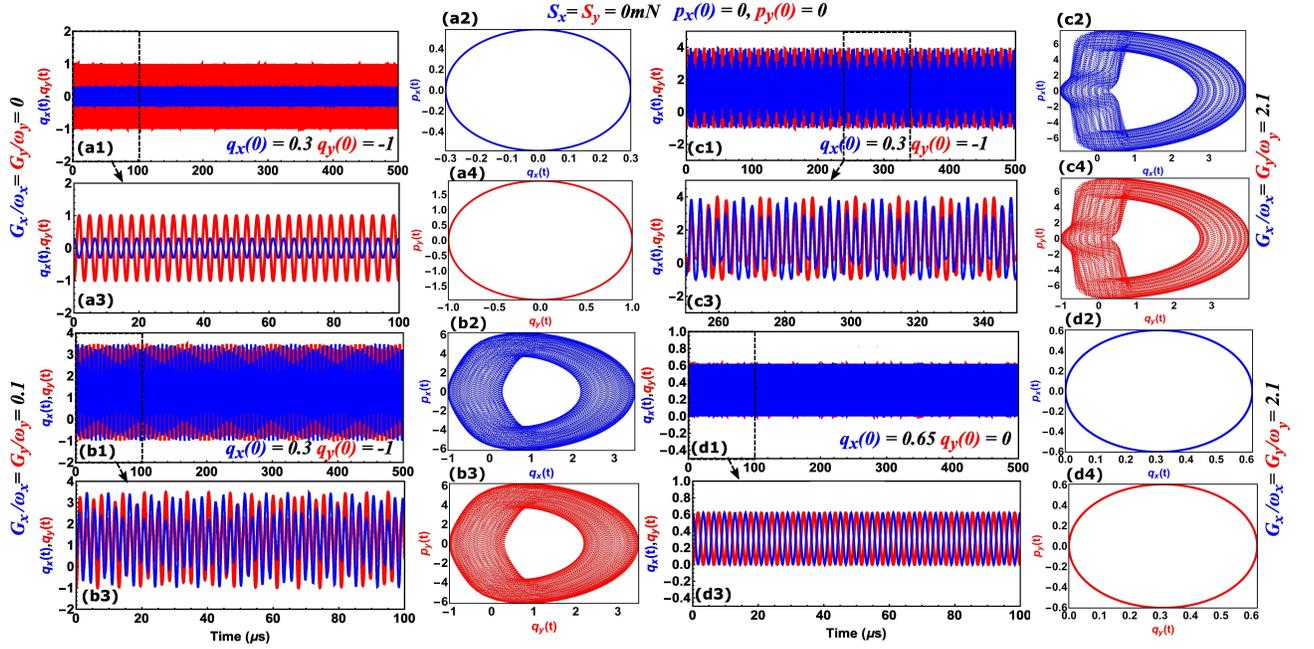}
	\caption{Spatio-temporal dynamics of the moving-end mirrors $q_x$ (blue) and $q_y$ (red) when $S_x=S_y=0$. (a), (b), (c) and (d) correspond to $G_x/\omega_x=G_y/\omega_y=0$, $0.1$, $2.1$ and $2.1$, respectively. ($\mathcal{O}1$) is the temporal response while ($\mathcal{O}3$) is its magnification, where $\mathcal{O}\rightarrow[a,b,c,d]$. ($\mathcal{O}2$) and ($\mathcal{O}4$) represent the corresponding phase space behavior for mirrors along $x$-axis and $y$-axis. For (a), (b) and (c), the initial conditions are $(q_x(0),p_x(0))=(0.3,0)$ and $(q_y(0),p_y(0))=(-1,0)$, while for (d), $q_x(0)=q_y(0)=0.65$ and $p_x(0)=p_y(0)=0$. Remaining parameters used are same as for Fig. \ref{fig2}.}
	\label{fig3}
\end{figure*}
However, when we exert the force of external mechanical drive (i.e. $S_x$ and $S_y$) on the mirrors, the regular and non-chaotic Poincar\'e sections start turning into mixed Poincar\'e sections containing chaotic seas and stable islands \cite{refc}, see Figs. \ref{fig2} (g-l). One can observe the significant interference between the phase space trajectories resulting in so much broadening of the circular pathways that they are indistinguishable from each other. This results in chaotic seas emerging all over the Poincar\'e sections but prominently appearing between $-2<q_{x,y}<0$ and $-3<p_{x,y}<3$. These chaotic signatures can be observed in both of the moving-end mirrors even when the mechanical drive is exerted only on the one moving-end mirror. It can be seen in Figs. \ref{fig2} (g) and \ref{fig2} (h) and Figs. \ref{fig2} (i) and \ref{fig2} (j), where $S_x=2mN$ and $S_y=0mN$, and $S_x=0mN$ and $S_y=2mN$, respectively. 

As both moving-end mirrors are indirectly coupled with each other through cavity field, the perturbations induced by spring effects of the cavity field (or mirror-cavity coupling) for any of the two mirrors also alter the dynamics of other moving-end mirror. Explicitly saying, the mechanical drive exerts force on the mirror which then transfers mechanical energy to the cavity field yielding to the unstable radiation pressure for other mirror. In this way, mechanical drive for any of the mirrors equally perturbs the motion of both mirrors. That is the reason why mirrors without mechanical drive possess similar Poincar\'e sections as mirrors with mechanical drives are illustrating. 

In presence of the mechanical drives for both moving-end mirrors, the Poincar\'e sections show similar patterns for each mirror, as illustrated in Figs. \ref{fig2} (k) and \ref{fig2} (l). Here the perturbation induced by both $S_x$ and $S_y$ enhances the interference between oscillatory pathways of phase space to such level that the mixed chaotic and stable features of Poincar\'e sections become more prominent. The stable islands -- the small circular structures -- correspond to such regions where if the moving-end mirror is initially located then it will remain stable and trapped in the stable region \cite{refc,refc1}. However, if any of the moving-end mirror is initially in chaotic region -- for example, appearing between $-2<q_{x,y}<0$ and $-3<p_{x,y}<3$ -- then it will be chaotically unstable and unpredictable. One can also note the emergence of stable islands in the form of  multi-fold, so-called, symmetric pattern originating for the center of Poincar\'e sections \cite{refc,refc1}. The multi-fold symmetric behavior is because of the cosine terms appearing in the effective Hamiltonian whose amplitudes basically define the strength of mechanical drive and are tuned over $\delta$ and $\phi$. These multi-fold symmetric mixed Poincar\'e sections are similar to the phase space dynamics studied in Ref. \cite{kashif1} and \cite{kashif2}, where the mechanical mirror of the cavity has been used as modulator for density excitation of ultra-cold atomic states and vise-versa. The mixed phase space behavior can further be explored but the emergence of chaos in Poincar\'e sections is enough to support the argument of our current study.   

\section{Spatio-Temporal Dynamics}\label{sec3}
\subsection{Rule of Mirror-Cavity Couplings}
The emergence of chaos in the dynamics of the moving-end mirrors can be further understood by measuring the spatio-temporal response. Here initial states, or conditions, of both of the mirrors play crucial rule, which we can carefully choose from the Poincar\'e sections. In a stable configuration -- where either mirrors are trapped in a stable island or they are isolated from the perturbed system -- both of the mirrors will follow stable and undamped osculations with time. It can be seen in Figs. \ref{fig3}(a1-a4), where both mirror-cavity couplings ($G_x/\omega_x=G_y/\omega_y=0$) as well as external mechanical force ($S_x=S_y=0$) is considered zero. Both of the mirror follow harmonic and stable (predictable) oscillations with time as can be seen in Figs. \ref{fig3}(a1) and \ref{fig3}(a3). If we plot their spatio-momentum phase space, for this particular configuration, both of the mirrors will follow single circular trajectory \cite{refc,refc1}, as illustrated in Figs. \ref{fig3}(a2) and \ref{fig3}(a4) for moving-end mirrors along $x$-axis and $y$-axis respectively. It should be noted that, for now, we have ignored the associated mechanical damping for both mirrors, i.e. $\gamma_x=\gamma_y=0$, in order to see the effects of mirror-cavity couplings and mechanical drive. However, later in the manuscript, we will discuss these effects. 

Further, one can note the difference in amplitudes of oscillations in both of the mirrors. It is because of the different initial condition, which we choose $(q_x(0),p_x(0))=(0.3,0)$ and $(q_y(0),p_y(0))=(-1,0)$ from the Poincar\'e sections illustrated in Fig. \ref{fig2}. As both mirrors are uncoupled to the cavity system, so their initial states will not be affected by the motion of each other, and they will oscillate with same initial amplitude. Although, different initial states in the isolated configuration are not showing much influence, except different oscillatory amplitudes. But, later, in presence of mirror-cavity coupling and external mechanical drive, not only the different amplitudes will be vanished, but these different initial states will also contribute to the engineering chaos.  

When we coupled the moving-end mirrors to the cavity field, both of the mirrors oscillate at same but $180^0$ shifted amplitudes, as can be seen in Figs. \ref{fig3}(b1-b4) and \ref{fig3}(c1-c4), where $G_x/\omega_x=G_y/\omega_y=0.1$ and $G_x/\omega_x=G_y/\omega_y=2.1$, respectively. It is because they both are now dependent on the vibrations of each other via cavity field. In other words, now cavity field exerts equal radiation pressure on both of the mirrors leading to the equal oscillatory amplitudes but with opposite phase. Further, because of their mutual coupling via cavity field, the amplitudes of both of the mirrors follow sinusoidal envelope.

\begin{figure}[htp]
	\includegraphics[width=8.5cm]{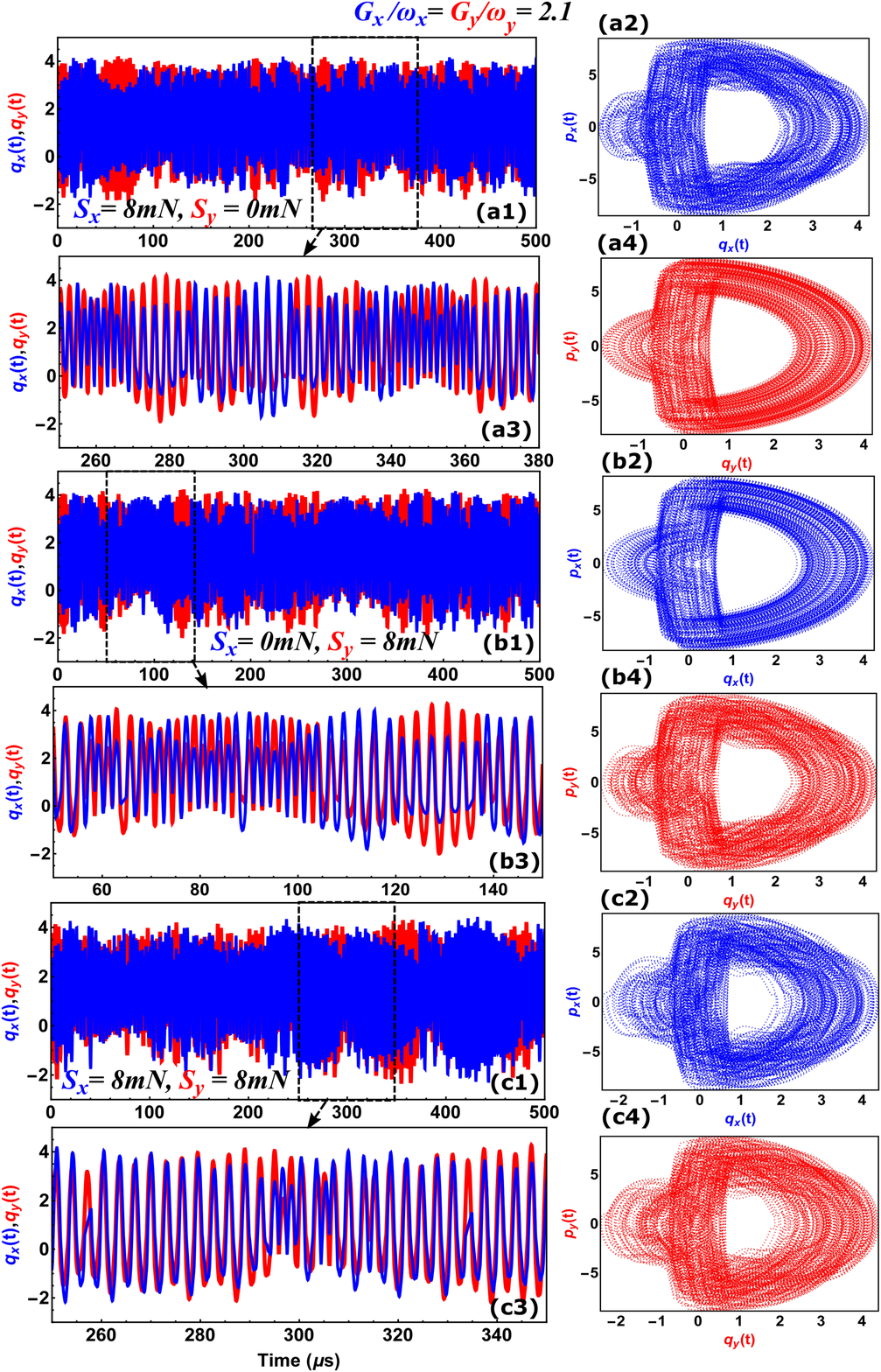}
	\caption{Spatio-temporal dynamics of the moving-end mirrors $q_x$ (blue) and $q_y$ (red) when $G_x/\omega_x=G_y/\omega_y=2.1$. Here (a), (b) and (c) are for $S_x=8 mN,S_y=0 mN$, $S_x=0 mN,S_y=8 mN$ and $S_x=8 mN,S_y=8 mN$, respectively. ($\mathcal{O}1$) represents the temporal response, with magnification ($\mathcal{O}3$). While ($\mathcal{O}2$) and ($\mathcal{O}4$) accommodate the relative phase spaces for the mirrors along $x$-axis and $y$-axis, where $\mathcal{O}\rightarrow[a,b,c]$. The initial conditions are considered as $(q_x(0),p_x(0))=(0.3,0)$ and $(q_y(0),p_y(0))=(-1,0)$. The other parameters used in calculation are same as mentioned in Fig. \ref{fig2}.}
	\label{fig4}
\end{figure}
Both of the moving-end mirrors follow significant nonlinear behavior with time, especially at $G_x/\omega_x=G_y/\omega_y=2.1$. The corresponding phase space plots also illustrate the effects of nonlinearities induced by mirror-cavity couplings, see Figs. \ref{fig3}(b2) and \ref{fig3}(c2) for ($q_x$,$p_x$) and Figs. \ref{fig3}(b4) and \ref{fig3}(c4) for ($q_y$,$p_y$). The trajectories in phase space take different and separated path in each temporal transition from the previous one, leading phase space to take oval disc shape. These nonlinear effects are more prominent at $G_x/\omega_x=G_y/\omega_y=2.1$, where a bump in oval disc shape appears around $p_{x,y}\rightarrow 0$, giving an impression of overlap between phase space trajectories. However, if we change the initial state of the moving-end mirrors by placing them in the stable region -- or in the small circle appearing around $p_{x,y}=0$ and $q_{x,y}=0.65$ in Figs. \ref{fig2}(e) and \ref{fig2}(f) -- then both of the mirrors will remain linearly trapped in these stable regions. It can be seen in Figs. \ref{fig3}(d1-d4), where initial conditions are considered as $(q_x(0),p_x(0))=(0.65,0)$ and $(q_y(0),p_y(0))=(0.65,0)$. Although, the mirror-cavity couplings bring significant amount of nonlinearity to the dynamics of moving-end mirrors, which could be enhanced with further increase in coupling strengths, but there are no sufficient signatures of chaos for the mirrors. 

\subsection{Mechanical Drive engineering Chaos}
In the case of external mechanical drive for both, or to any, of the mirrors, both of the moving-end mirrors show unstable and unpredictable temporal response, as can be seen in Fig. \ref{fig4}. The amplitudes of oscillations, for both mirrors, now randomly vary with time without following any patterns. These random oscillatory amplitudes then yield into overlap and interference between phase space trajectories. Like Poincar\'e sections, both mirrors possess chaotic signatures even when the mechanical drive is exerted only on one of the mirrors. However, unlike the Poincar\'e sections, the temporal dynamics for both cases (when mechanical drive is present for one mirror and absent for other mirror and when mechanical drive is present for the other mirror and absent for the first mirror) are different for mirror with mechanical drive from mirror without mechanical drive. It can be observed in Figs. \ref{fig4}(a1) and \ref{fig4}(a3), Figs. \ref{fig4}(b1) and \ref{fig4}(b3), where $S_x=8 mN,S_y=0 mN$, and $S_x=0 mN,S_y=8 mN$, respectively. But, apparently, phase spaces for the moving-end mirrors without mechanical drive are showing similar patterns (see Figs. \ref{fig4}(a4) and \ref{fig4}(b2)) and are also same for the moving-end mirrors with mechanical drives (see Figs. \ref{fig4}(a2) and \ref{fig4}(b4)). It is because, the phase space trajectories corresponding to each temporal recurrence interfere with the previous one creating a complex phase space structure from which it is not easy to extract the temporal evolution \cite{refc,refc1}. 

\begin{figure}[tp]
	\includegraphics[width=8.5cm]{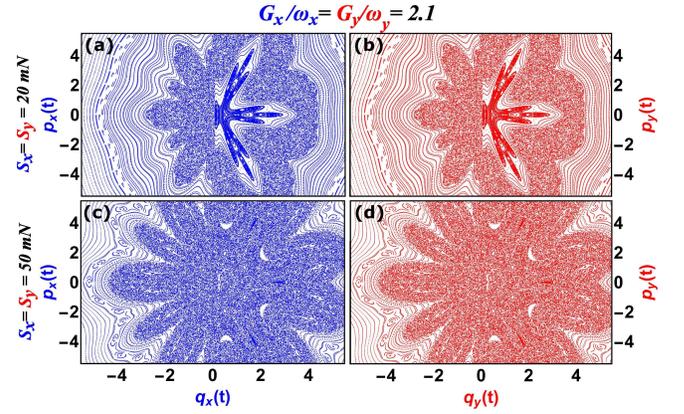}
	\caption{Poincar\'e surface of section for both of the moving-end mirrors $q_x,p_x$ (blue) and $q_y,p_y$ (red) oriented along $x$-axis and $y$-axis, at $G_x/\omega_x=G_y/\omega_y=2.1$. Here (a) and (b) correspond to the strengths of mechanical drives $S_x=S_y=20 mN$, while (c) and (d) illustrate the effects of mechanical drives when their strengths are $S_x=S_y=20 mN$. Remaining configuration and parameters are same as used in Fig. \ref{fig2}.}
	\label{fig5}
\end{figure}
The interference between phase space trajectories for the moving-end mirrors with mechanical drives is much prominent as compared to the phase space of mirrors without mechanical drive. It reveals that the mechanical drive induces more perturbation effects to the interacting mirror, apparently, unlike the Poincar\'e sections. In fact, in the Poincar\'e sections, we take temporal elapse for a specific time, which leads to the suppression of such differences. However, if we exert equal force of external mechanical drives to both of the moving-end mirrors then they both will show similar behavior, as can be seen in Figs. \ref{fig4}(c1-c4). One can observe the emergence of significant chaotic patterns not only in temporal dynamics but also in their corresponding phase spaces. The distance between phase space trajectory decreases with each recurrence resulting in interference between trajectory pathways. Such interference is the signature for the occurrence of chaos. 
\begin{figure}[tp]
	\includegraphics[width=8.5cm]{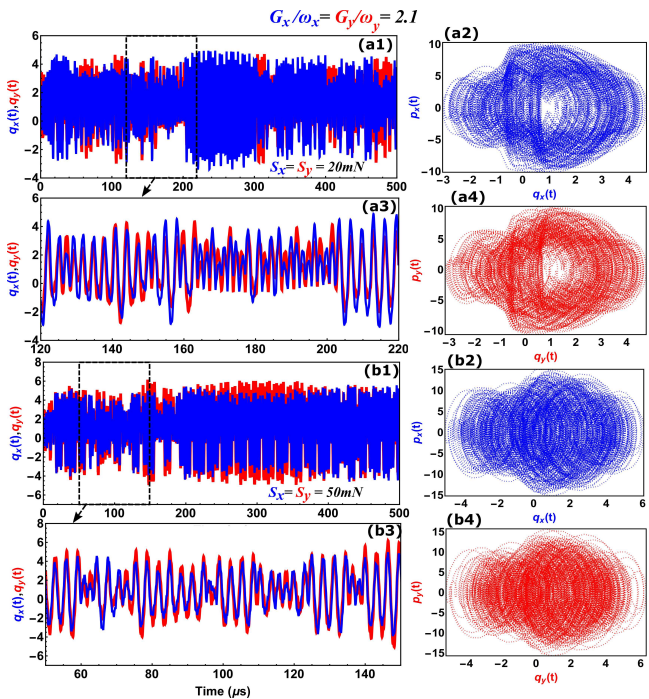}
	\caption{Spatio-temporal response of the moving-end mirrors $q_x$ (blue) and $q_y$ (red) at mirror-cavity coupling when $G_x/\omega_x=G_y/\omega_y=2.1$. (a) and (b) illustrate the effects of strength of mechanical drive $S_x=S_y=20 mN$ and $S_x=S_y=50 mN$, respectively. Here ($\mathcal{O}1$) contains the temporal dynamics with magnification ($\mathcal{O}3$), with $\mathcal{O}\rightarrow[a,b]$. While ($\mathcal{O}2$) and ($\mathcal{O}4$) are the corresponding phase spaces for mirrors along $x$-axis and $y$-axis. The initial states of the mirrors are considered as $(q_x(0),p_x(0))=(0.3,0)$ and $(q_y(0),p_y(0))=(-1,0)$. The remaining parameters used in numerical calculation are same as for Fig. \ref{fig2}.}
	\label{fig6}
\end{figure}
\begin{figure*}[htp]
	\includegraphics[width=14cm]{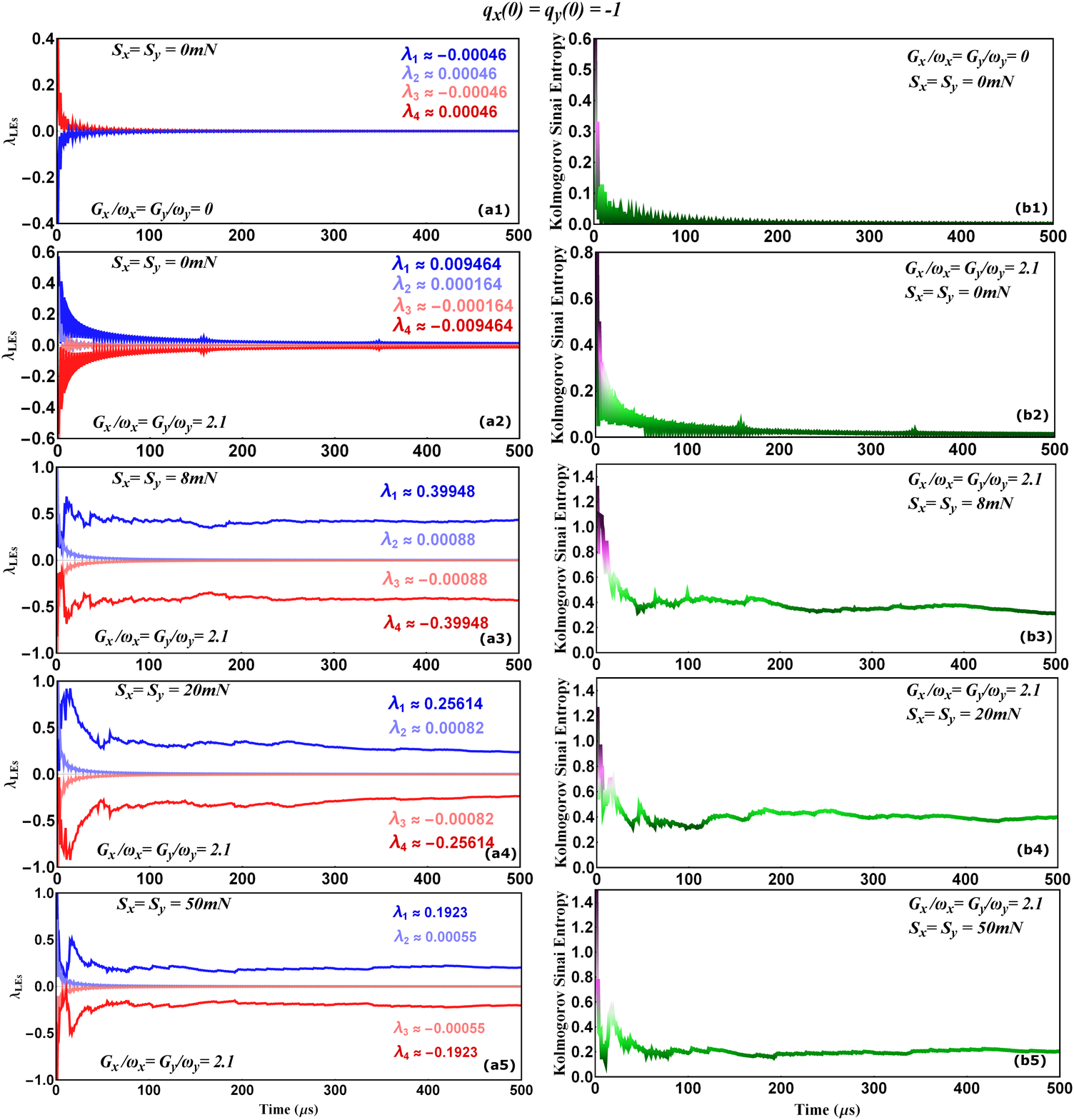}
	\caption{Quantitative analysis for the occurrence of Chaos. (a1-a5) illustrate the Lyapunov exponent spectrum while (b1-b5) show the Kolmogorov-Sinai Entropy. For ($\mathcal{O}1$-$5$), the coupling is $G_x/\omega_x=G_y/\omega_y=0$, $2.1$, $2.1, 2.1$, and the strength of external drive is $S_x=S_y=0, 0, 8, 20, 50 mN$, respectively, where $\mathcal{O}\rightarrow[a,b]$. The initial conditions for these calculations are $(q_x(0),p_x(0))=(q_y(0),p_y(0))=(-1,0)$. Further, $\lambda_{1-4}$ correspond to the long-term saturation values of the Lyapunov exponents. Remaining parameters used are same as for Fig. \ref{fig2}.}
	\label{fig7}
\end{figure*}

\begin{figure*}[htp]
	\includegraphics[width=14cm]{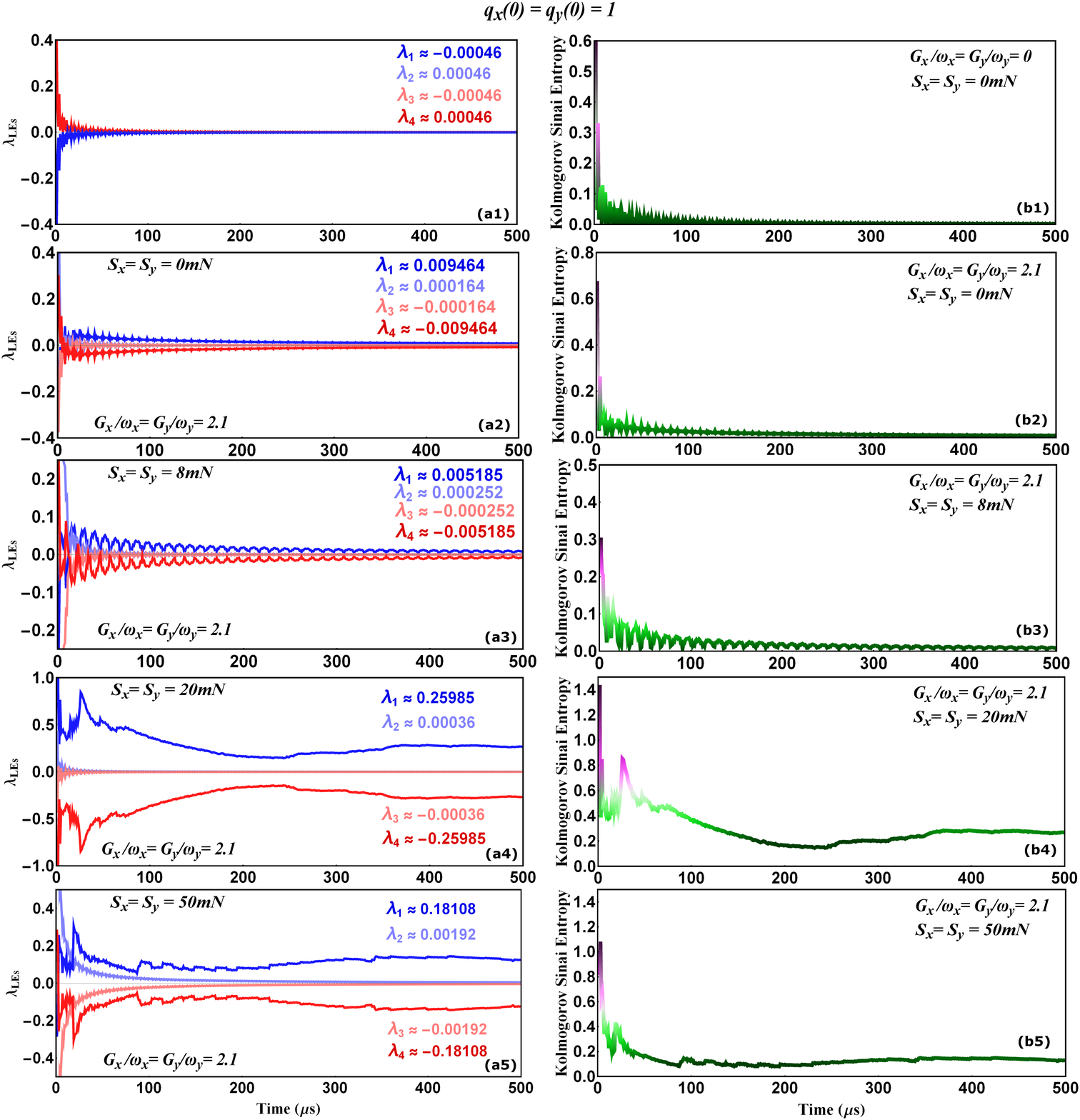}
	\caption{Similarly as in Fig.\ref{fig7}, Here (a1-a5) illustrate the Lyapunov exponent spectrum while (b1-b5) show the Kolmogorov-Sinai Entropy. For ($\mathcal{O}1$-$5$), the mirror-cavity coupling is $G_x/\omega_x=G_y/\omega_y=0$, $2.1$, $2.1, 2.1$, while the external drive strength is $S_x=S_y=0, 0, 8, 20, 50 mN$, respectively, where $\mathcal{O}\rightarrow[a,b]$. The initial conditions for these calculations are $(q_x(0),p_x(0))=(q_y(0),p_y(0))=(1,0)$. Further, $\lambda_{1-4}$ correspond to the long-term saturation values of the Lyapunov exponents. Remaining parameters used are same as for Fig. \ref{fig2}.}
	\label{fig8}
\end{figure*}

Here, we considered the same initial state for both of the moving-end mirrors as we considered for previous discussion, i.e. $(q_x(0),p_x(0))=(0.3,0)$ and $(q_y(0),p_y(0))=(-1,0)$ for mirror along $x$-axis and $y$-axis, respectively. If we recall Fig. \ref{fig2}(k) and \ref{fig2}(l) -- which correspond to the same parameters as we considered in Figs. \ref{fig4}(c1-c4) -- and compare these initial conditions, one can easily note that the mirror along $x$-axis is initially located in stable (or substable) region. While the mirror along $y$-axis is initially located in the chaotic region on the left-half plane of Poincar\'e sections. Conventionally, as the mirror oriented along $x$-axis initially lies in the stable region of Poincar\'e sections, so it should illustrate stable temporal dynamics within that stable region \cite{refc,refc1}. But it is not happening. Not only the mirror located along $y$-axis, which indeed initially is in chaotic region, is demonstrating chaotic behavior but the mirror along $x$-axis is also showing similar chaotic dynamics. The reason of this is the cavity field mediated coupling between both of the moving-end mirrors. Whenever any of the mirrors is possessing unstable and chaotic features, it will transfer these features to the other mirror as well. Thus, in order to obtain stability, both of the mirrors should be initially in stable region. If any of the mirrors is initially not in stable island, then both of the moving-end mirrors will illustrate chaos. 

If we further increase the magnitude of external mechanical drive, the amount of perturbation to the spatio-temporal dynamics gets further strong. This will then lead to the enhanced interference between phase space trajectories and increased chaotic region in Poincar\'e sections, as can be seen in Fig. \ref{fig5}(a) and \ref{fig5}(b), and Fig. \ref{fig5}(c) and \ref{fig5}(d), where the strengths of external mechanical drives are increased to $S_x=S_y=20 mN$ and $S_x=S_y=50 mN$, respectively. The remaining parametric configuration is same as used for Fig. \ref{fig2}. One can note that the chaotic seas now occupy the majority of the space in Poincar\'e sections. In the case of $S_x=S_y=20 mN$, there are some stable islands emerging around the center on the right-half plane. But these stable regions get significantly shrunk in the case of $S_x=S_y=50 mN$. This is the case where chaos completely dominates the system. 

\begin{figure}[tp]
	\includegraphics[width=8.5cm]{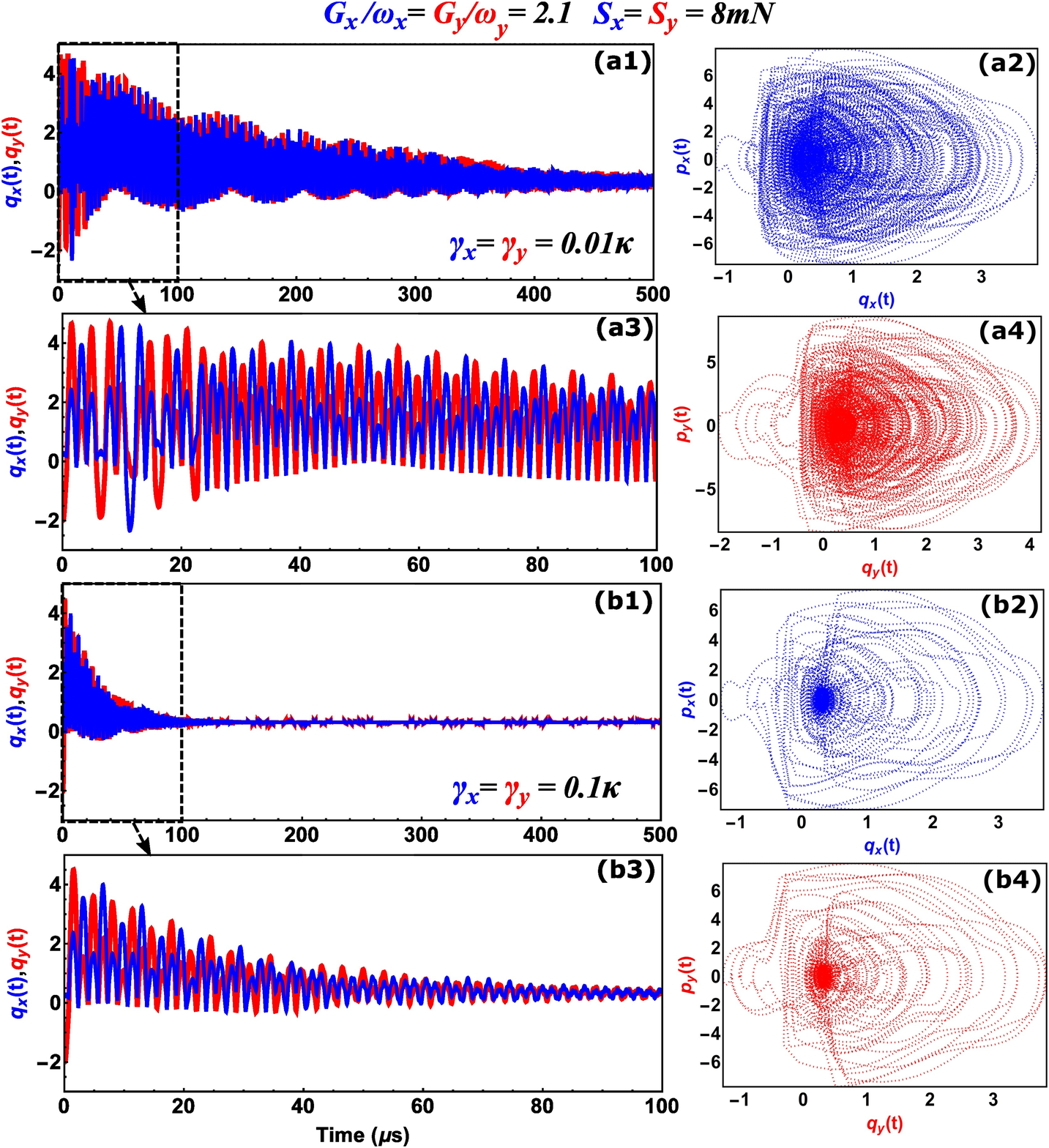}
	\caption{The effects of mechanical damping on the spatio-temporal response of moving-end mirrors $q_x$ (blue) and $q_y$ (red) when $G_x/\omega_x=G_y/\omega_y=2.1$ and $S_x=S_y=8 mN$. (a) and (b) accommodate the mechanical damping rate $\gamma_x=\gamma_y=0.01\kappa$ and $\gamma_x=\gamma_y=0.1\kappa$, respectively, associated with the mechanical motion of the mirrors. Similar to the previous temporal responses, here ($\mathcal{O}1$) corresponds to the temporal dynamics with its magnification ($\mathcal{O}3$) and ($\mathcal{O}2$) and ($\mathcal{O}4$) are the corresponding phase spaces for mirrors along $x$-axis and $y$-axis, here $\mathcal{O}\rightarrow[a,b]$. The initial states for both of the mirrors are $(q_x(0),p_x(0))=(0.3,0)$ and $(q_y(0),p_y(0))=(-1,0)$. While the remaining parameters are same as for Fig. \ref{fig2}.}
	\label{fig9}
\end{figure}
Similar enhanced chaotic effects can be observed if we measure the temporal response of both of the mirrors with same system configuration, as illustrated in Fig. \ref{fig6}(a1-a4) and \ref{fig6}(b1-b4), where $S_x=S_y=20 mN$ and $S_x=S_y=50 mN$, respectively. The temporal response now gets further random (unpredictable) and chaotic, but it is not easy to extract these enhancements from temporal dynamics itself. However, corresponding phase space plots show significant enhancements with increased external mechanical drives. The perturbation induced overlap and intersections between recurring phase space trajectories now overtake the maximum of the mirror dynamics yielding to chaos. Especially, in the case of $S_x=S_y=50 mN$, where the chaos apparently occupies the whole space destroying any possibility of pattern occurring in phase space \cite{refc,refc1}. Thus, these results and discussion reveal the chaotic dynamics for both of the moving-end mirrors induced by external mechanical drives.     

\section{Quantitative Analysis: Lyapunov Exponent to Kolmogorov-Sinai Entropy}\label{sec4}
Previous discussion of the manuscript contains well-enough qualitative analysis of the occurrence of chaos in the system. But it is also important to have an idea about the quantitative behavior of the chaos. For this purpose, we calculated all possible Lyapunov Exponent \cite{lyp1,lyp2,lyp3} and collected Kolmogorov-Sinai entropy \cite{KSL1,KSL2} of the system. Lyapunov exponent is a parameter that measures infinitesimally small separation between phase space trajectories. That rate of separation can be different for different initially oriented phase space trajectory vectors, especially in a Hamiltonian system. These separation rates determine a spectrum of Lyapunov exponents proportional to the dimensions of associated degrees of freedom \cite{lyp1,lyp2,lyp3}. As the largest valued Lyapunov exponent among the spectrum accounts for the largest separation of phase space trajectories, therefore, it represents the occurrence of chaos in the system. 

In our system, we used Gram-Schmidt numerical process to orthonormalize the vectors in the Jacobian matrix equation, which is obtained from the set of QLEs of the system \cite{lyp1,lyp2,lyp3,KSL1,KSL2}. The eigenvalues for the solution of Jacobian equation basically defined the separation between phase space trajectories in the form of Lyapunov exponent spectrum, as illustrated in Fig.\ref{fig7}(a1-a5) and Fig.\ref{fig8}(a1-a5). One thing should be noted here, these Lyapunov exponents represents collective response of all associated subsystems or degrees of freedom with the system. Secondly, the number of Lyapunov exponents in the spectrum should be double of the number of subsystems associated with the system, in accordance with Oseledets theorem imposed on Jacobian matrix equation \cite{lyp1,lyp2,lyp3,KSL1,KSL2}. The largest Lyapunov exponent in the spectrum will defines the amount of chaos occurring in the system at specific time depending on the chosen parameters and initial conditions.   

To further extend our quantitative analysis of chaos or collective disorderness happening in the system, we calculate the Kolmogorov-Sinai entropy of the system. Kolmogorov-Sinai entropy illustrates the amount of total disorderness of any system and, conventionally by using Pesin’s identity (Pesin’s entropy formula) \cite{KSL1,KSL2}, can be calculated by adding all positive Lyapunov exponents of the system \cite{KSL3,KSL4,KSL41},
  \begin{eqnarray}
  	\mathcal{H}_{KS}&\approx& \sum_{\lambda_{LE}^i>0}^{2\mathbb{N}}\lambda_{LE}^i. \label{LE} 
  \end{eqnarray}
Here, $\lambda_{LE}^i$ corresponds to the Lyapunov exponent associated with $i^{th}$ degree of freedom and $\mathbb{N}$ is the total number of associated degrees of freedom. The sum of all positive Lyapunov exponents basically provides the upper bound for the entropy. But to calculate highest possible disorderness, one can approximate total summation of positive Lyapunov exponents equation to entropy \cite{KSL1,KSL2}. 

In Fig.\ref{fig7}, (a1)-(a5) illustrate the Lyapunov exponent spectrum of the system and (b1)-(b5) illustrate the corresponding Kolmogorov-Sinai entropy. The initial conditions are $(q_x(0),p_x(0))=(q_y(0),p_y(0))=(-1,0)$, chosen from the Poincar\'e surface of sections shown in Fig.\ref{fig2}. In Figs.\ref{fig7}(a1) and \ref{fig7}(b1), where $G_x/\omega_x=G_y/\omega_y=0$ and $S_x=S_y=0$, one can note that both Lyapunov exponent spectrum $\lambda_{LEs}$ ($\lambda_{1-4}$) and Kolmogorov-Sinai entropy saturate to almost zero with time. It means that in absence of external mechanical drives and mirror-cavity couplings, the system possesses no chaos and behaves linearly. There are some initial fluctuations depending upon chosen initial conditions. But these fluctuations appeared to be zero with time. In presence of mirror-cavity coupling, the lyapunov spectrum and Kolmogorov-Sinai entropy appear to be again saturated near zero but these values and fluctuations are slightly higher than the zero coupling case, as illustrated in Figs.\ref{fig7}(a2) and \ref{fig7}(b2), where  $G_x/\omega_x=G_y/\omega_y=2.1$ and $S_x=S_y=0$. It is because of the nonlinearities occurring because of the mirror-cavity coupling. These nonlinear signatures can also be seen in Poincar\'e surface of sections shown in Figs.\ref{fig2}(e) and \ref{fig2}(f) at $(q_x(0),p_x(0))=(q_y(0),p_y(0))=(-1,0)$. However, as discussed in the case of phase space and temporal response, there are no sufficient signatures for the occurrence of chaos. 

However, when we exert external mechanical drive, the behavior of Lyapunov spectrum gets completely changed. Now the largest Lyapunov exponent, after initially fluctuating with time, saturates to a positive value around $\lambda_1\approx0.39948$, as illustrated in Fig.\ref{fig7}(a3), where $S_x=S_y=8mN$. While second and third Lyapunov exponents saturate to near zero $\lambda_2\approx0.00088$ and $\lambda_3\approx-0.00088$ with time, respectively. The fourth Lyapunov exponent possesses the lowest possible value $\lambda_4\approx-0.39948$, which appears to be symmetric with largest Lyapunov exponent. Here the positiveness of largest Lyapunov exponent represents the occurrence of the chaos and its value defines the quantitative occurrence of the chaos. The corresponding Kolmogorov-Sinai entropy is shown in Fig. \ref{fig7}(b3), where one can note the nonzero saturation of entropy with time. Similar to the largest Lyapunov, it indicates the quantitative amount of chaos in the system and, as previously said, the Kolmogorov-Sinai entropy defines the upper bound to largest Lyapunov exponent (or largest Lyapumov defines lower bound for Kolmogorov-Sinai entropy), its value will always be greater than the largest Lyapunov exponent. These both indicators sufficiently and quantitatively prove the occurrence of chaos in the system. 

If we further increase the force of external drives, both Lyapunov exponent and Kolmogorov-Sinai entropy remains positive, as can be seen in Figs. \ref{fig7}(a4) and \ref{fig7}(b4), Figs. \ref{fig7}(a5) and \ref{fig7}(b5), where $S_x=S_y=20mN$ and $=20mN$, respectively. However, one can note that the saturated values are slightly less than the case of $S_x=S_y=8mN$. It means that the amount of disorder at higher strengths of external drives is less as compared to the lower values of external drives. It is because of initial conditions. Means, at particular initial values, the amount of disorder can be higher at lower perturbations as compared to the disorder at higher perturbations. It can be well understood by visualizing Poincar\'e surface of sections. If the mirrors are initially located at such point where the Poincar\'e surface of sections contains more chaos at low strengths of external drive as compared to the high values, then obviously Lyapumov exponents will also have higher values at low external force. The occurrence of mixed Poincar\'e surface of sections with more stable island could limit or modify the amount of disorder in the system. To further illustrate that we plotted Lapunov exponents and Kolmogorov-Sinai entropy shown in Fig. \ref{fig7} at different initial condition, as can be seen in Fig. \ref{fig8}, where $(q_x(0),p_x(0))=(q_y(0),p_y(0))=(1,0)$.  

At $G_x/\omega_x=G_y/\omega_y=0$ and $S_x=S_y=0$ (Figs. \ref{fig8}(a1) and \ref{fig8}(b1)), both Lyapumov and Kolmogorov-Sinai entropy saturate to zero with time similarly like the previous case. At $G_x/\omega_x=G_y/\omega_y=2.1$ and $S_x=S_y=0$ (Figs. \ref{fig8}(a2) and \ref{fig8}(b2)), again both show slightly different but generally similar behavior with previous case. But when we increase the strength of external force to $S_x=S_y=8mN$, both Lyapumov and Kolmogorov-Sinai entropy saturate very low values as compared with the previous case of initial conditions $(q_x(0),p_x(0))=(q_y(0),p_y(0))=(-1,0)$ and show, sort-of, oscillatory behavior with time. However, when we increase the external force, both of these factors move to higher values and show different behavior with the previous case, see Figs. \ref{fig8}(a4) and \ref{fig8}(b4), Figs. \ref{fig8}(a5) and \ref{fig8}(b5), where $S_x=S_y=20mN$ and $=20mN$, respectively. 

This comparison between Lyapumov spectrum and Kolmogorov-Sinai entropy at different initial conditions clearly states that these both factors crucially depend on initial configuration of the system. One can also conclude that although Lyaponov exponents give quantitative value of chaos but at particular initial condition, not overall. On the other hand, Poincar\'e surface of sections, as illustrated in Fig. \ref{fig2}, not only give the idea of occurrence of chaos but also illustrates the dynamical behavior of that chaos over all possible initial conditions. The demonstration of Kolmogorov-Sinai entropy in this section also enhances the understanding of disorderness and this procedure could be used to connect disorderness and chaos with multiple phenomenon, like quantum entanglement and localization. 

\section{Effects of Mechanical damping}\label{sec5}
In the manuscript, so far, we haven't considered the effects of mechanical damping rates ($\gamma_{x,y}$), associated with the oscillatory motion of the moving-end mirror, on the spatio-temporal dynamics. The purpose of doing so is to clearly illustrate the chaotic dynamics mediated by external mechanical drive, which would not be much clear in presence of mechanical damping. However, in this section, we intend to include the effects of associated mechanical damping rates in presence of external drive $S_{x,y}$. As expected, the mechanical damping rates damps (limits) the oscillatory amplitudes of both mirrors, depending upon the magnitude of damping \cite{refc,refc1}. This will result in continuous squeeze of the spatio-temporal evolution, as illustrated in Figs. \ref{fig9} (a1,a3) and Figs. \ref{fig9} (b1,b3), where damping rates are considered as $\gamma_x=\gamma_y=0.01\kappa$ and $\gamma_x=\gamma_y=0.1\kappa$, respectively. One can note that when we consider the higher magnitudes of mechanical damping, the squeeze effects get significantly enhanced yielding into a, sort-of, saturated motion along the origin, i.e. $q_{x,y}\approx0$, as can be seen in the case of $\gamma_x=\gamma_y=0.1\kappa$. 

However, the chaotic behavior remains there during the saturation process, even in the saturated state, as can be observed in Figs. \ref{fig9} (a2,a3) and Figs. \ref{fig9} (b2,b3), where the phase spaces corresponding to the relative temporal response are illustrated. The phase space trajectories are appeared to be randomly saturating towards the center or the origin like their temporal counter parts. During this saturation process and even in the saturated states, phase space trajectories notably overlap and interact with each other yielding to the chaos. These trajectories interference, in the present configuration, are more prominent in the case of $\gamma_x=\gamma_y=0.01\kappa$. But, for $\gamma_x=\gamma_y=0.1\kappa$, the interference effects are there but most of the trajectories are themselves shifted to the saturated domain. One can further extract the underlying features induced by the mechanical damping in spatio-temporal dynamics by further investigating and considering higher damping rates. But here we are focused to the main theme of the study, which is the emergence of chaos with external mechanical drives in our setup.        

\section{Conclusion}\label{sec6}
In conclusion, we investigate chaos in a four-mirror optomechanical cavity induced by external mechanical drives. A strong pump laser longitudinally (along $x$-axis) drives the cavity building up a strong cavity mode. That cavity mode, after getting split from beam splitter, then interacts with the transversely located (along $x$-axis and $y$-axis) moving-end mirrors. The radiation pressure, exerted by the cavity mode, results in oscillatory motion for both of the mirrors. Two transverse laser beams externally interact with each of the moving-end mirrors, resulting in external mechanical drives that perturb their oscillatory motion. By constructing the Poincar\'e surface of sections over a wide interval of initial conditions, we illustrate the spatio-momentum dynamics of both of the moving-end mirrors under the influence of mirror-cavity coupling and external mechanical drive. We find that the presence of mechanical drive turns the regular and stable Poincar\'e sections to the mixed Poincar\'e sections, containing both stable islands and chaotic seas. To further enhance the understanding for the occurrence of chaos, we illustrate the spatio-temporal dynamics and plot the corresponding phase spaces. The presence of the mechanical drives yields not only into unpredictable temporal evolution for both moving-end mirrors, but it also engineers interference between the recurrence trajectories of phase spaces leading to chaos. These chaotic features get significantly enhanced with the increased magnitude of mechanical drive. 

In order to quantitatively analyze the chaos, we calculated the Lyapumov spectrum of exponents and Kolmogorov-Sinai entropy. We find that the increase in the force of external drives induces more amount of disorderness in the system but that disorderness is crucially dependent on initial conditions of the system. The calculation of Kolmogorov-Sinai entropy with the Lyapumov spectrum of exponents also provides another way to related disorderness and chaos with disorder mediated phenomenon like quantum entanglement and dynamical localization. We further illustrate the effects of mechanical damping associated with moving-end mirrors and show that the damping rates squeeze the oscillatory amplitudes of temporal dynamics, but these chaotic effects are still present in the evolution of the moving-end mirrors. The findings of our work are crucial in order to construct hybrid quantum-classical picture of complex dynamical systems and could provide a platform to test chaos induced quantum mechanical phenomenon.    
     
\begin{acknowledgments}
	K.A.Y. acknowledges the support of Research Fund for International Young Scientists by NSFC under grant No. KYZ04Y22050, Zhejiang Normal University research funding under grant No. ZC304021914 and Zhejiang province postdoctoral research project under grant number ZC304021952. G.X.L. acknowledges the support of National Natural Science Foundation of China under Grant Nos. 11835011 and 11774316.
\end{acknowledgments}

\end{document}